# Development of an inverse identification method for identifying constitutive parameters by metaheuristic optimization algorithm: Application to hyperelastic materials

G. Bastos[1], A. Tayeb[1], N. Di Cesare[2], J.-B. Le Cam[1], E. Robin[1]

[1] Univ. Rennes 1, CNRS, IPR (Institut de Physique de Rennes), UMR 6251, F-35000 Rennes, France

[2] Université Bretagne Sud, UMR CNRS 6027, IRDL, F-56100 Lorient, France

**ABSTRACT**

In the present study, a numerical method based on a metaheuristic parametric algorithm has been developed to identify the constitutive parameters of hyperelastic models, by using FE simulations and full kinematic field measurements. The full kinematic field is measured at the surface of a cruciform specimen submitted to equibiaxial tension. The sample is reconstructed by FE to obtain the numerical kinematic field to be compared with the experimental one. The constitutive parameters used in the numerical model are then modified through the optimization process, for the numerical kinematic field to fit with the experimental one. The cost function is then formulated as the minimization of the difference between these two kinematic fields. The optimization algorithm is an adaptation of the Particle Swarm Optimization algorithm, based on the PageRank algorithm used by the famous search engine Google.

**INTRODUCTION**

The constitutive parameters of hyperelastic models are generally identified from three homogeneous tests, basically the uniaxial tension, the pure shear and the equibiaxial tension. From about 10 years, an alternative methodology has been developed [1, 2, 3, 4], and consists in performing only one heterogeneous test as long as the field is sufficiently heterogeneous. This is typically the case when a multiaxial loading is applied to a 3 branch or a 4-branch cruciform specimen, which induces a large number of mechanical states at the specimen surface. The induced heterogeneity is generally analysed through the distribution of the biaxiality ratio and the maximal eigen value of the strain. The Digital Image Correlation (DIC) technique is generally used to retrieve the different mechanical states induces, and provides the full kinematic field at the specimen surface, i.e. a large number of experimental data to be analysed to identify the constitutive parameters of the behaviour model considered.

Several methods have been recently developed to identify parameters from experimental field measurements, typically the finite elemnt updating method (FEMU), the constitutive equation gap (CEGM), the virtual fields method (VFM), the equilibrium gap method (EGM) and the reciprocity gap method (RGM). These method are fully reviewed in [5] and have been shown to be adequate for problems with moderate numbers of unknown constitutive parameters.

In the present study, a metaheuristic parametric algorithm is proposed to identify the constitutive parameters of a behaviour model that actually minimize the cost function in the FEMU approach. The optimization algorithm used in both based on the Particle Swarm Optimization (PSO) algorithm and the artificial smart PageRank algorithm used by the famous search engine Google. This algorithm allows the minimization of the full kinematic field differences by modifying the constitutive parameters, while minimizing the CPU calculation time. Even though the final objective is the identification of complex constitutive models, the Mooney's model [6] is presented in this paper.

## EXPERIMENTAL SETUP

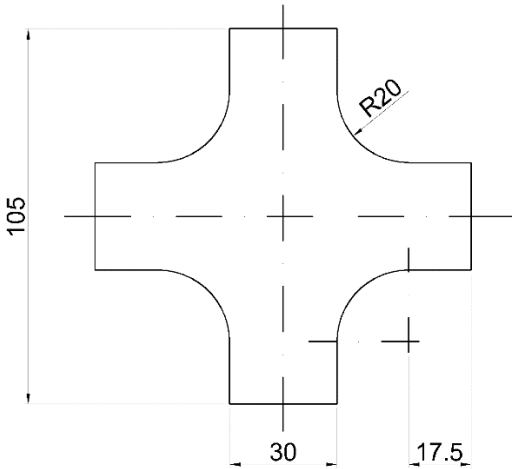

Figure 1: Specimen geometry (dimensions in mm).

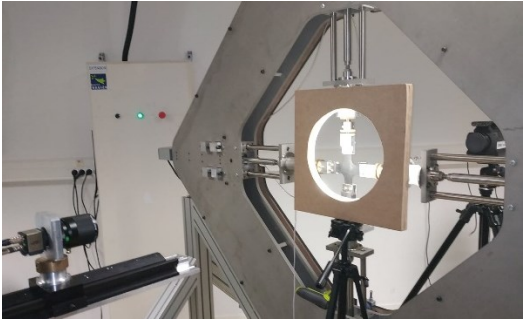

Figure 2: Home-made biaxial testing machine

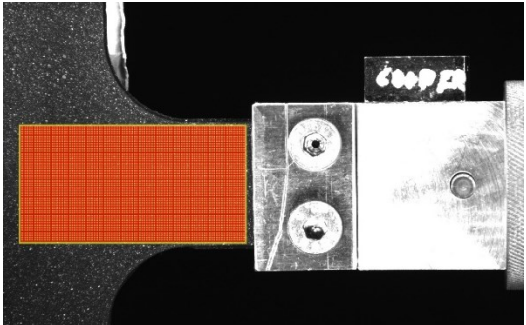

Figure 3: Zone Of Interest for the DIC technique

The material use dis a carbon black filled natural rubber. The specimen geometry is shown in Figure 1. It is a 105mm long and 2 mm thick cruciform specimen. Figure 2 presents an overview of the experimental setup composed of a home-made biaxial testing machine and an optical camera. The machine is composed of four independant RCP4-RA6C-I-56P-4-300-P3-M (IAI) electrical actuators. They were driven by a PCON-CA-56P-I-PLP-2-0 controller and four PCON-CA (IAI) position controllers. The actuators were controlled by an in-house LabVIEW program. Two cell loads, whose capacity is equal to 1094 N, store the force variation in the two perpendicular directions. In the present study, one equibiaxial loading was carried out in a way that the specimen's centre was motionless for the displacement measurement to be easier. The displacement and loading rate were set at 70 mm and 150 mm/min respectively for the four independent actuators.

Images of the specimen surface at increasing stretches were stored at a frequency equal to 5hz with a IDS camera equipped with a 55 mm telecentric objective. The charge-coupled device (CCD) of the camera has 1920 × 1200 joined pixels. The Digital Image Correlation (DIC) technique is used to determine the displacement field at the sample surface. It consists in correlating the grey levels – a white paint is sprayed on the sample to improve the image contrast - between two different images of a given zone, each image corresponding to different strain levels. The software used for the correlation process was SeptD [7], and a uniform cold lightning was ensured by a home-made LED lamp. The spatial resolution, defined as the smallest distance between two independent point was equal to 4 pixels corresponding to 0.34268 mm. The ZOI used to make the digital correlation of the displacement field is represented in Figure 3.

## NUMERICAL MODEL

A finite element calculation is performed by assuming plane stress state and material incompressibility. The four-node PLANE182 ANSYS element is used. The mesh is made of 9600 nodes, and 9353 elements. A biaxial traction load is obtained by prescribing the same displacement of 70mm on the four branched of the sample. The two-parameters hyperelastic Mooney model is used for the calculation. The values of the constitutive parameters C01 and C10 are changing at each iteration of the optimization process, as described in the next section. The value of the incompressibility parameter has been set to 1E-5 Mpa$^{-1}$ for all the calculations proceeded.

As the FE discretization of the numerical model and the used camera precision are not the same, it could be tricky to compare the two kinematic fields. To overcome that problem, the experimental kinematic field is fitted by a polynome-based function. In this way, the numerical kinematic field will be compared, for each node, with the experimental field in the exact same position of the sample. As it can be tricky to fit an experimental field by a polynome-based function, it has been decided to fit the experimental field by as many functions as the FE model contains nodes. The used function is given as a 5 order polynome in the X and Y directions, so 21 nodes and their respective cooerdinates and field values are needed to te able to retrieve the polynome's coefficients. So, for every node in the FE model, the 21 closest nodes are used to retrieve the corresponding kinematic polynome. By using this method, the difference between the experimental field and the polynome-based function is less than 0.2 mm for every point.

## METAHEURISTIC OPTIMIZATION STRATEGY

The aim of the optimization process is here to find the constitutive parameters for the numerical kinematic field to fit the experimental one. Because of the linear relation between C01 and C10, the displacement can be the same for two different pairs of constitutive parameters. The force has then to be fitted too. The cost function is then defined as the squared difference between the experimental and the numerical fields, considering the force too, as follows:

$$\min \sum_{i=1}^{N} \frac{1}{N} \left(\frac{U_{x,exp} - U_{x,num}}{U_{x,exp}}\right)^2 + \frac{F_{exp} - F_{num}}{F_{exp}}$$

Where N is the number of nodes in the numerical Zone of Interest, $U_{x,exp}$ is the polynome-based experimental displacement, $U_{x,num}$ is the numerical horizontal displacement, $F_{exp}$ is the experimental horizontal force, and $F_{num}$ the numerical horizontal force.

The optimization algorithm used is an adaptation of the classical Particle Swarm Optimization algorithm. In this version, all the particles are influenced by all the others, by considering this influence to be adapted as a function of the respective performance of the particles. The population of PSO particles is then considered as a Markov chain, in which the particles are the nodes, and the transition probabilities between them are the links between them. For each particle, the PageRank value – that is the steady state of the considered Makov chain – is given by the following equation (2). In this way, the PageRank value of each particle is deduced from its performance compared to the best one.

$$\pi_{target}^T = \left| \frac{fitness(G_{best}) \times 100}{fitness(G_{best}) - fitness(P) + \varepsilon} \right|$$

It it then possible to deduce the transition connectivity matrix **C** giving the influence of all the particles on all the others by using a pseudo-random process. The classical equations of PSO are then modified, weighing the influence of all the particles by using the components of **C**, as follow:

$$V_i^{t+1} = \omega \times V_i^t + c_1 \times rand_1 \times (P_{i,best}^{t+1} - X_i^t) + c_2 \times rand_2 \times \sum_{j=1}^{n} C_{ij} \times (P_{j,best}^{t+1} - X_i^t)$$

$$X_i^{t+1} = X_i^t + V_i^{t+1}$$

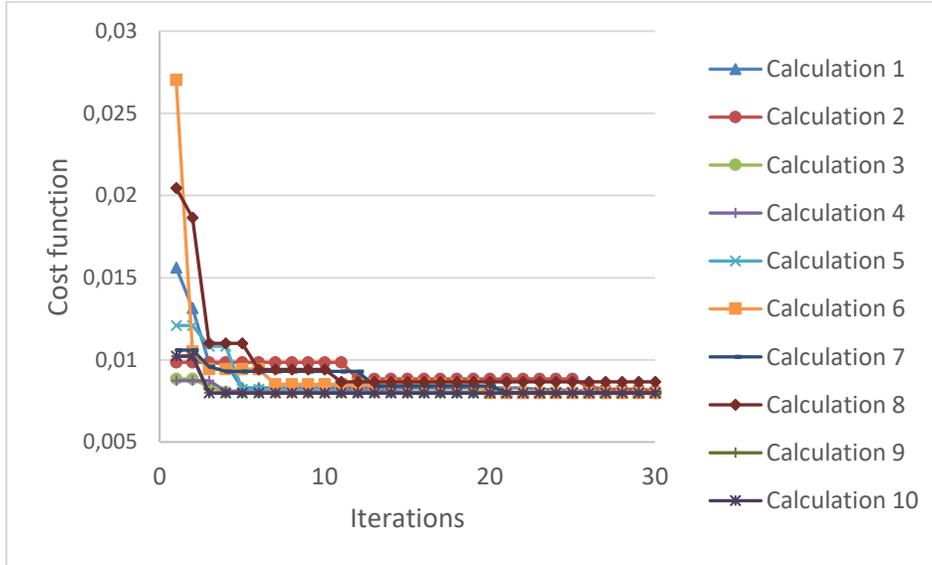

Figure 4: Convergence curves on the 10 optimization calculation launched

Where $V_i^{t+1}$ is the speed of the $i^{th}$ particle at iteration $t+1$, $c_1$ and $c_2$ are confident parameters, $\omega$ is the inertia weight, $X_i^{t+1}$ is the position of particle $i$ at iteration $t+1$, $rand_1$ and $rand_2$ are random numbers in [0,1], $P_{j,best}^{t+1}$ is the personal best position of particle $i$ at iteration $t+1$, and **C** is the transition connectivity matrix of the considered Markov chain. This Inverse-PageRank-PSO algorithm is fully described in [8].

## RESULTS AND DISCUSSION

As the particles are initially randomly def ined, the optimization has been launched 10 times, to compare the obtained solutions, and be sure that the global minimum of the cost function has been reached. The convergence curves of the 10 launched optimization calculation are represent in in Figure 4. The obtained values of the cost function and design variables are given in Table 1.

The validation of the optimized results is checked by comparing the experimental displacements and efforts in the sample with the numerical optimized one. In the final numerical model, the values of C01 and C10 have been set to the mean of the obtained optimized values found in the 10 different calculations launched. Figure 5 shows the difference between the experimental polynome-based kinematic field, and the optimized numerical one, for every point in the ZOI.

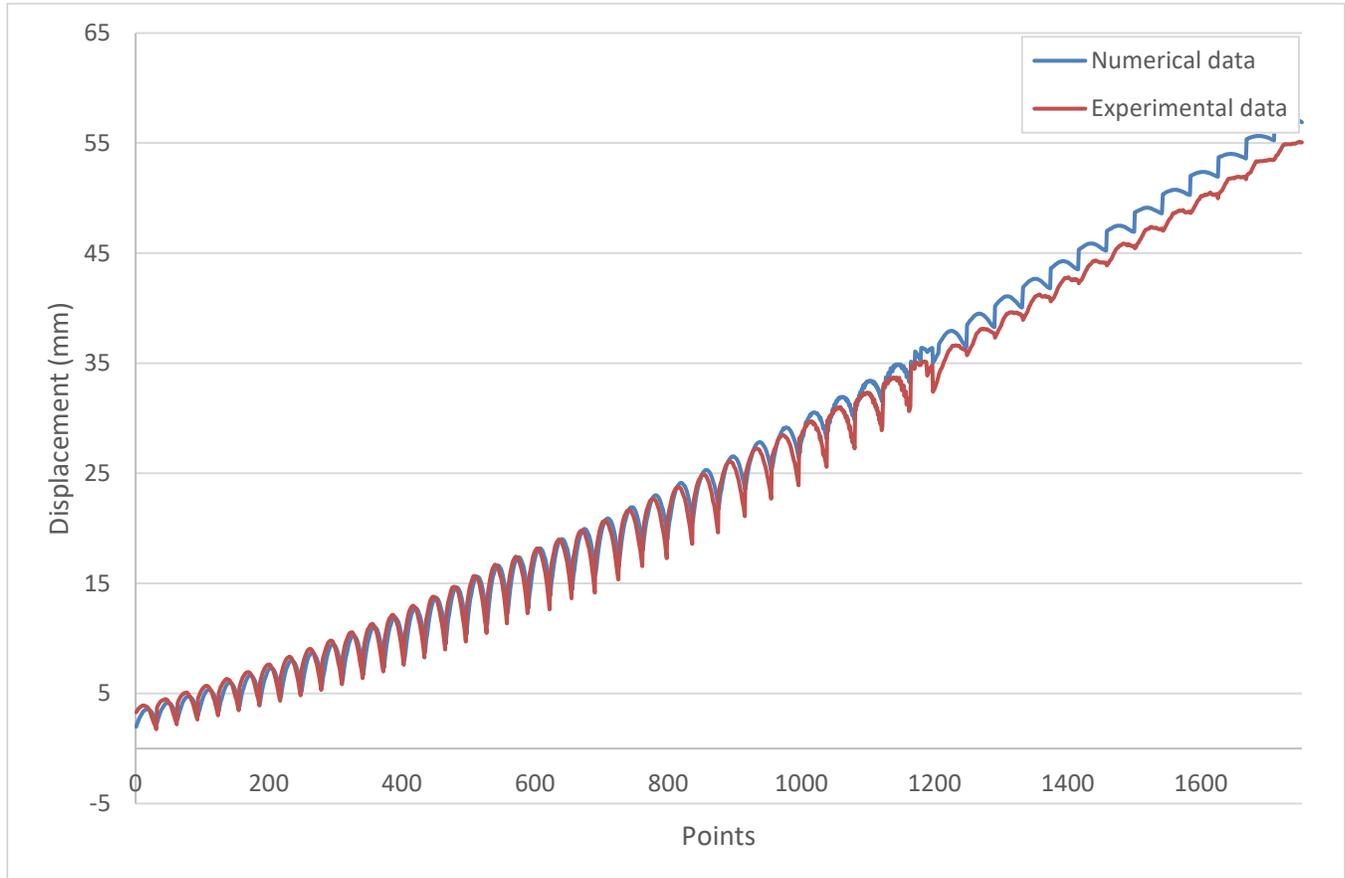

Figure 5: Comparison between the kinematic fields after the optimization process

| Obtained results | | | |
|---|---|---|---|
| *Optimization number* | **Cost function** | **C01** | **C10** |
| 1 | 8.05E-3 | 0.5159 | 0.01792 |
| 2 | 8.13E-3 | 0.5116 | 0.022541 |
| 3 | 7.98E-3 | 0.5188 | 0.020369 |
| 4 | 7.98E-3 | 0.5197 | 0.018879 |
| 5 | 7.98E-3 | 0.5176 | 0.019914 |
| 6 | 7.99E-3 | 0.5175 | 0.020986 |
| 7 | 8.01E-3 | 0.5153 | 0.02008 |
| 8 | 8.03E-3 | 0.5146 | 0.022146 |
| 9 | 8.15E-3 | 0.5109 | 0.021573 |
| 10 | 7.98E-3 | 0.5202 | 0.019215 |
| *Mean* | 8.03E-3 | 5.16E-1 | 2.03E-2 |
| *Std* | 6.475E-5 | 3.19E-3 | 1.15E-4 |

Table 1: Obtained results

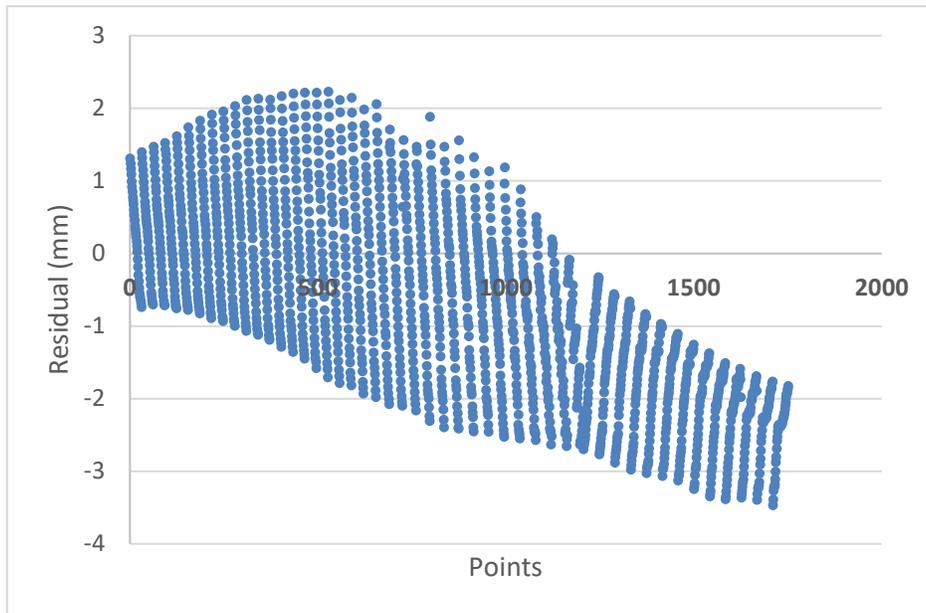

This difference is presented in a quantitative way in Figure 6 showing the difference between the two fields for every point of the Zone of Interest. One can note that the difference is always less than 6.5% of the experimental kinematic field. Considering the force, the experimental value was 176.02N, while the numerical value obtained with the optimized values of $C_{01}$ and $C_{10}$ is 176.37N, which leads to a difference up to 0.4%.

Figure 6: Difference between the kinematic fields for every point of the ZOI

## CONCLUSIONS

This work is proposing a new inverse identification method based on the coupling of experimental kinematic fields retrieved by DIC, and the using of a PSO-based parametric optimization algorithm. Experimental and numerical kinematic fields are compared to finally be fitted through the optimization process, while the constitutive parameters and smartly modified. Applied on a Mooney model, this process is able to find the constitutive parameters reproducing the mechanical response of the sample, while minimizing the number of optimization iterations. The constitutive parameters found by the optimization process are actually giving a numerical model that retrieves precisely the entire kinematic experimental field.

## ACKNOWLEDGEMENTS
The authors thank the Regoin Bretagne, Rennes Métropole and Université Bretagne Loire for their financial support. Cooper Standard is aknoledged for providing the specimens. Authors also thank Dr. Mathieu Miroir, M. Vincent Burgaud, and M. Mickael Le Fur for having designed the biaxial testing machine, Gilles Marckmann and Prof. Michel Grédiac for the fruitfull discussions.